\documentstyle[epic,eepic,12pt]{ioplppt}
\def\intR{\int_{\bf R}}              
\def\E{\mbox{\bf E}}                  
\def\r{\mbox{\bf r}}                  
\def\nr{\mbox{{\bf :}}}               
\def\<{\langle}                       
\def\>{\rangle}                       
\def\beq{\begin{eqnarray}}            
\def\enq{\end{eqnarray}\noindent}     
\begin{document}
\title{Two-photocurrent devices}
\author{Matteo G. A. Paris}
\address{Arbeitsgruppe 'Nichtklassiche Strahlung' der Max-Planck-Gesellschaft \\
         Rudower Chaussee 5, 12489 Berlin,Germany \\
         {\tt PARIS@PHOTON.FTA-BERLIN.DE}}
\begin{abstract}
Heterodyne, eight-port homodyne and six-port homodyne detectors belong to 
the class of two-photocurrent devices. Their full equivalence in probing 
radiation field has been proved both for ideal and not fully efficient 
photodetectors. The output probability distribution has been also 
evaluated for a generic probe mode.
\end{abstract}
\section{Introduction}
\label{s:intro} 
In order to gain information about a quantum state of the radiation field
one has to measure some observable. The measurement process unavoidably 
involves some kind of interaction, thus coupling the mode under examination
to one or more other modes of the field. One has to admit that, in general, 
the measured observable is not defined on the sole Hilbert space of the signal
mode \cite{bus}. On the contrary, it reflects properties of the global state, 
possibly entangled, coming from the interaction among the signal mode and the 
set of the probe modes. Sometimes, it is possible to get rid of these probe 
modes, so that the statistics of the output is related only to the quantum 
statistics of the signal mode. This, as an example, is the case of balanced 
homodyne detectors \cite{cha}. In that case, in fact, an appropriate rescaling
of the output photocurrent allows to completely neglect the local oscillator 
in the definition of the measured observable, which simply results to be 
proportional to the signal field quadrature
$\hat a (\phi )=\frac{1}{2}(a^{\dag}e^{i\phi}+ae^{-i\phi})$.
\par
However, in more general cases this procedure cannot be pursued 
and the output statistics always remind us the way we are probing 
the signal under examination. This is a common feature of three relevant 
detection schemes in quantum optics, which provide different setup for jointly 
measuring a couple of photocurrents. They are the heterodyne \cite{ysh,sha,sho},
the eight-port homodyne \cite{wa1,wa2,wa3,lai} and the recently introduced
six-port homodyne detectors \cite{our,zuc}.\par
It is a purpose of this paper to show that, although they involve very
different ways of coupling signal to probe modes, they provide the same
information on the signal under examination. Actually, we will prove 
their full equivalence by demonstrating that their output photocurrents
have the same operatorial form. More surprisingly, this remains true also
when taking into account the inefficiency of photodetection, even though the 
latter occurs at very different stages in the three detection schemes. \par
The paper will be organized as follows. In the next section we briefly
review how to describe inefficient photodetection in terms of beam
splitters and ideal detectors. In the three sections
\ref{s:eig},\ref{s:het} and \ref{s:tri} we will examine the three
detection schemes respectively, in order to show that all of them jointly 
measure the real and the imaginary part of the complex photocurrent 
\cite{sha,two}
\begin{eqnarray}
\widehat {\cal Z} = a + b^{\dag}
\label{photo}\:,
\end{eqnarray}
being $a$ the signal mode and $b$ a probe mode. The inefficiency of 
realistic photodetectors will be also taken into account, in order to show 
it does not affect the equivalence of the considered detection schemes. \par
In section ref{s:pom} we will derive the output statistics of considered 
schemes as members of the more general class of two-photocurrent devices. 
Section \ref{s:out} will close the paper with some 
concluding remarks.
\section{Inefficient photodetection}
\label{s:eta}
The final stage of any detection scheme is represented by
photodetection, namely counting photon through their conversion
to electric pulses. Let us consider a light beam $\hat\rho$ entering
a phototube which converts to electric pulses a fraction $\zeta$ of
the incoming photons. By keeping open the detector window for a time
interval $T$, the probability $P_m (T)$ of counting $m$ photons is
given by \cite{kel}
\begin{eqnarray}
P_m (T) = \hbox{Tr}\left\{\hat\rho\nr{{[\zeta\hat I (T)T]^m }\over{m!}}
\exp [-\zeta\hat I (T)T]\nr\right\}        
\label{pc-gen}\;,
\end{eqnarray}
where $\nr\:\:\nr$ denotes normal ordering of operator and
$\hat I (T)$ is the beam intensity
\begin{eqnarray}
\hat I (T)={2\epsilon_0 c \over T}\int_0^T \hat{\E}^{(-)} (\r,t)\cdot
\hat{\E}^{(+)} (\r,t) dt
\label{intensity}\;.
\end{eqnarray}
$\hat{\E}^{(\pm)} (\r,t)$ denotes the positive (negative) frequency
part of the field. For a single-mode field excited in a stationary state
Eq. \ref{pc-gen} can be rewritten as
\begin{eqnarray}
P_m^{\eta} = \hbox{Tr}\left\{\hat\rho\:{{(\eta a^{\dag} a)^m} \over{m!}}\exp
(-\eta a^{\dag} a)\:\right\} 
\label{pc-singlemode}\;,
\end{eqnarray}
being $[a,a^{\dag}]=1$ the single mode field operator and
$\eta=\zeta c\hbar\omega/V$ the global quantum efficiency of the
photodetectors.  For unit quantum efficiency this coincides with the
actual photon number distribution of the state under examination
\begin{eqnarray}
P_m^1 = \rho_{mm} \equiv \langle m|\hat\rho | m \rangle
\;,\label{Pn}
\end{eqnarray}
whereas, in the realistic case of non-unit quantum efficiency Eq. 
(\ref{pc-singlemode}) becomes a binomial convolution \cite{man}.
In formula
\begin{eqnarray}
P_m^{\eta} =\sum_{n=m}^{\infty} \rho_{nn}
\left(\begin{array}{c} n\\m\end{array}\right) \eta^m (1-\eta )^{n-m}
\label{conv_n}\;.
\end{eqnarray}
Let us now consider the scheme in Fig. \ref{f:eta}. The signal mode $a$
is impinged in a beam splitter of transmissivity $\tau$ whose second
port $b$ is placed in the vacuum. A perfect ($\eta=1$) photodetection on
the exiting mode reveals $m$ photons with a probability
\begin{eqnarray}
P_m = \hbox{Tr} \left(\hat U_{\tau}(\hat\rho\otimes|0\>\< 0|)\hat
U^{\dag}_{\tau}|m\>\<m|\otimes\hat 1  \right)
\label{conv_n1}\;,
\end{eqnarray}
being $\hat 1$ the identity operator on the second input of the beam
splitter and
$$\hat U_{\tau}= \exp\left\{i
\arctan \sqrt{\frac{1-\tau}{\tau} \left(a^{\dag} b - ab^{\dag}\right)}
\right\}\:,$$
the evolution operator of the beam splitter.
A straightforward calculation shows that
\begin{eqnarray}
P_m =  \sum_{n=m}^{\infty}\rho_{nn}
\left(\begin{array}{c} n\\m\end{array}\right) (1-\tau)^{n-m} \tau^m
\label{conv_n2}\;.
\end{eqnarray}
Eq. (\ref{conv_n2}) coincides with Eq. (\ref{conv_n}) for
$\tau=\eta$. This means that a photodetection process with efficiency
$\eta$ is equivalent to a perfect photodetection process performed on the
original signal mixed with vacuum by a beam splitter with a value of the
transmissivity equal to the quantum efficiency \cite{rea,opt}.
In the following we will adopt this equivalent scheme.
\section{Eight-port homodyne detector}
\label{s:eig}
Eight-port homodyne detector is known for a long time for the joint
determination of phase and amplitude of the field in the microwave
domain. It was introduced in the optical domain by Walker
and Carrol \cite{wa1} and successively analyzed by different authors
\cite{wa2,wa3,lai,ma1,ma2,ma3,ban,fre,leo,cos,rip,hen,qsm,dmr}. 
A schematic diagram of the
experimental setup is reported in Fig. \ref{f:eig}.
There are four balanced beam splitters whereas a $\pi/2$ phase shifter
is inserted in one arm.
The four input modes are denoted by $a_k$, $k=1,...,4$ whereas the detected
output modes are denoted by $b_k$, $k=1,...,4$. There are four identical
photodetectors whose quantum efficiency is given by $\eta$.
The {\em noise} modes used to take into account inefficiency, according to
the scheme of the previous section, are denoted by $u_i$, $i=1,...,4$.
We consider $a_1$ as the  signal mode, whereas $a_2$ is referred to be
the idler of the device. The mode $a_3$ is unexcited, whereas $a_4$ is
placed in a highly excited coherent state $|z \rangle$ provided by an
intense laser beam (local oscillator). The detected photocurrents are
$\hat I_k=b^{\dag}_k b_k$, $k=1,...,4$
which form the eight-port homodyne observables
\begin{eqnarray}
\widehat {\cal Z}_1 = \frac{\hat I_2 - \hat I_1}{2\eta |z|} \nonumber \\
\widehat {\cal Z}_2 = \frac{\hat I_4 - \hat I_3}{2\eta |z|}
\label{8phot}\:.
\end{eqnarray}
The latter are derived by rescaling the difference photocurrent, each
of them obtained in an homodyne scheme. For this reason eight-port
homodyne is known also as double-homodyne detection.
In Eq. (\ref{8phot}) $\eta$ denotes the quantum efficiency of the
photodetectors whereas $|z|$ is the intensity of the local oscillator.
In order to obtain $\widehat {\cal Z}_1 $  and $\widehat {\cal Z}_2$
in terms of the input modes we first note that the input-output mode
transformation is necessarily linear, as only passive components are
involved in the detection scheme \cite{st1,st2}. Thus, we can write
\begin{eqnarray}
b_k =  \sum_{l=1}^{4} M_{kl} a_l
\label{linear}\:,
\end{eqnarray}
where the transformation matrix can be computed starting
from the corresponding transformations for the beam splitters
and the phase shifter \cite{zei}
\begin{eqnarray}
{\bf M} = \frac{1}{\sqrt{4}}
\left[
\begin{array}{cccc}
1 &  1 &  1 &   1 \\
1 &  i & -1 &  -i \\
1 & -1 &  i &  -1 \\
1 & -i & -1 &   i
\end{array}
\right]
\label{matrix}\:.
\end{eqnarray}
Eqs. (\ref{linear}) and (\ref{matrix}) together with the equivalent
scheme for the inefficient detector leads to the following expression
for the output modes
\begin{eqnarray}
b_1 =&\sqrt{\eta}\;\left[a_1+ a_2 + a_3+ a_4\right]&+\sqrt{1-\eta}\; u_1  
\nonumber \\
b_2 =&\sqrt{\eta}\;\left[a_1+ ia_2 -a_3-i a_4\right]&+\sqrt{1-\eta}\; u_2 
\nonumber  \\
b_3 =&\sqrt{\eta}\;\left[a_1- a_2 + ia_3- a_4\right]&+\sqrt{1-\eta}\; u_3 
\nonumber \\
b_4 =&\sqrt{\eta}\;\left[a_1- ia_2 -a_3+ i a_4\right]&+\sqrt{1-\eta}\; u_4
\label{output}\:.
\end{eqnarray}
Upon inserting Eqs. (\ref{output}) in Eq. (\ref{8phot}) and considering
the limit of highly excited local oscillator we obtain the
eight-port photocurrents in terms of the input modes
\begin{eqnarray}
\fl \widehat {\cal Z}_1&=&
\hat a_{1} ( 0) + \hat a_{2}( 0 )
+ \sqrt{\frac{1-\eta}{\eta}} \left[\hat u_{1} (0 ) -
\hat u_{2} (0)\right] + O[\frac{1}{|z|}]
\nonumber \\
\fl \widehat {\cal Z}_2&=&
-\hat a_{1} (\pi/2) + \hat a_{2} (\pi/2)
+ \sqrt{\frac{1-\eta}{\eta}} \left[ \hat u_{3} (\pi/2) - \hat u_{4} (\pi/2)
\right] + O[\frac{1}{|z|}]
\label{zphot}\:.
\end{eqnarray}
In Eq. (\ref{zphot}) $\hat a(\phi )$ denotes a quadrature of the field.
The complex photocurrent ${\cal Z}={\cal Z}_1+i{\cal Z}_2$
is given by
\begin{eqnarray}
{\cal Z}=a_1 + a_2^{\dag}
\label{8zed}\:,
\end{eqnarray}
for unit quantum efficiency, whereas for non unit quantum efficiency 
it becomes a 
Gaussian convolution of Eq. (\ref{8zed}), we will consider this point 
in detail in \ref{s:pom}.
\par
It is worth noticing here that the mode transformation defined by Eqs.
(\ref{linear}) and (\ref{matrix}) is distinctive for a canonical
$4\times 4$-port linear coupler as defined in Refs. \cite{ig1,ig2}. It has
been rigorously shown \cite{zei} that a $N\times N$-port linear coupler
can always be realized in terms of a number of beam splitters and
phase-shifters. However, this implementation is, in general,  not
unique. The interest of eight-port homodyne scheme 
lies in the fact it provides the minimal scheme for realizing a  
$4\times 4$-multiport \cite{min}.
If one considers the multiport as a given {\em black-box}  device the
eight-port homodyne scheme can be depicted as in Fig. \ref{f:rough}.
This will facilitate the comparison with the six-port homodyne detection
presented in Section \ref{s:tri}.
\section{Heterodyne detector}
\label{s:het}
Heterodyne detection scheme is known for a long time in radiophysics.
It has been introduced in the domain of optics \cite{ysh,sha,sho,y82} 
in order to
describe the joint measurement of two conjugated quadratures of the field.
The term 'heterodyne' is used as the involved field modes are excited on
different frequencies. \par
In Fig. \ref{f:het} we show a schematic diagram of the detector.
We denote by $\hat{\E}_S$ the signal field, whereas $\hat{\E}_{LO}$
describes the local oscillator. The field $\hat{\E}_L$ accounts for the
losses due to inefficient photodetection.
The input signal is excited in a single mode (say $a$) at the frequency
$\omega$, whereas also the local oscillator is excited in only one mode
at the frequency $\omega_0$. This local oscillator mode is placed in
a strong coherent state $|z \rangle$ by means of an intense laser beam.
The beam splitter has a transmissivity given by $\tau$, whereas the
photodetectors shows quantum efficiency $\eta$.
The heterodyne output photocurrents are given by the real $\widehat{\cal Z}_1$
and the imaginary $\widehat{\cal Z}_2$ part of the complex photocurrent
$\widehat{\cal Z}$. The latter is obtained after the rescaling of the output
photocurrent $\hat I$, which is  measured at the intermediate frequency
$\omega_I = \omega -\omega_0$.
By Fourier transform of Eq. (\ref{intensity}) we have
\begin{eqnarray}
\hat I (\omega_I )= \intR d\omega ' \; \hat{\E}^{(-)}_O (\omega ' + \omega_I  )
\:\hat{\E}^{(+)}_O ( \omega ')
\;,\label{hetero1}
\end{eqnarray}
being $\hat{\E}^{(\pm)}_O$ positive and negative part of the output
field. In terms of the input fields Eq. (\ref{hetero1}) can be written as
\begin{eqnarray}
\fl \hat I (\omega_I ) =  \intR \!\!\!d\omega '
&&\left[\sqrt{\eta\tau} \hat{\E}^{(-)}_S (\omega ' + \omega_I  ) +
\sqrt{\eta (1-\tau)}\hat{\E}^{(-)}_{LO} (\omega ' + \omega_I  ) +
\sqrt{1-\eta}\hat{\E}^{(-)}_L (\omega ' + \omega_I  ) \right]
\nonumber \\ &&
\left[\sqrt{\eta\tau} \hat{\E}^{(+)}_S (\omega ' ) +
\sqrt{\eta (1-\tau)}\hat{\E}^{(+)}_{LO} (\omega ' ) +
\sqrt{1-\eta}\hat{\E}^{(+)}_L (\omega ' ) \right] 
\;.\label{hetero2}
\end{eqnarray}
Heterodyne photocurrent is obtained by the following rescaling
\begin{eqnarray}
\widehat{\cal Z} =\lim_{\tau \rightarrow 1, |z| \rightarrow \infty}
\frac{\hat I (\omega_I )}{ |z| \eta \sqrt{\tau(1-\tau)}} \qquad \;
\mbox{with $|z| \sqrt{1-\tau}$ constant}
\;. \label{hetero3}
\end{eqnarray}
Physically this definition corresponds to require a very intense local
oscillator, which however is allowed only for a little mixing with the
signal mode \cite{par}. In this limit only terms containing the local oscillator
field $\E^{(\pm )}_{LO} (\omega_0 )$ at the frequency $\omega_0$ can
survive in Eq. (\ref{hetero2}), so that we have
\begin{eqnarray}
\widehat{\cal Z} =  \widehat{\cal Z}_1 + i \widehat{\cal Z}_2
\;,\label{hetero4}
\end{eqnarray}
where
\begin{eqnarray}
\widehat {\cal Z}_1&=& \hat a(0) + \hat c (0) + \sqrt{\frac{1-\eta}{\eta}}\:
\left(\hat u_{1} (0) - \hat u_{2} (0) \right)
\nonumber \\
\widehat {\cal Z}_2&=& - \hat a (\pi/2) + \hat c (\pi/2)+ 
\sqrt{\frac{1-\eta}{\eta}}\:
\left( \hat u_{1} (\pi/2) -  \hat u_{2}(\pi/2)
\right)\;.\label{hetero5}
\end{eqnarray}
In writing Eq. (\ref{hetero5}) we have substituted
\begin{eqnarray}\begin{tabular}{l}
$ c \leftarrow \hat{\E}^{(+)}_S (2\omega_0-\omega ) $\\    \\
$u_1 \leftarrow \hat{\E}^{(+)}_L (\omega ) $\\             \\
$u_2 \leftarrow \hat{\E}^{(+)}_L (2\omega_0-\omega )$
\end{tabular}\:, \label{hetero6}
\end{eqnarray}
for the relevant modes involved.
Provided  $u_1$ and $u_2$ to be noise modes placed
in the vacuum the expression (\ref{hetero5}) for the heterodyne
photocurrents leads to the identical output statistics of eight-port 
homodyne photocurrents, the mode $c$ playing the role of the idler of 
the device. The full equivalence of the two detection schemes has been 
thus proved.
\section{Six-port homodyne detector}
\label{s:tri}
A linear, symmetric three-port optical coupler is a straightforward
generalization of the customary lossless symmetric beam splitter.
The three input modes $a_i$, $i=1,2,3$ are combined to form 3 output
modes $b_j$, $j=1,2,3$. In analogy to lossless beam splitters, which
are described by unitary 2$\times$2 matrices~\cite{cam},
any lossless triple coupler is characterized by a unitary
$3\times 3$ matrix~\cite{zap,zin}. For the symmetric case we have the form
\par\noindent\begin{eqnarray}
{\bf T} = \frac{1}{\sqrt{3}}\left(\begin{array}{ccc}
1 & 1 & 1 \\  &   &   \\ 1 & \exp\{i\frac{2\pi}{3}\} &
\exp\{-i\frac{2\pi}{3}\} \\ & & \\1 & \exp\{-i\frac{2\pi}{3}\}
& \exp\{i\frac{2\pi}{3}\}\end{array}\right)\label{T3M}\;,
\end{eqnarray}\par\noindent
where each matrix element $T_{ij}$ represents the transmission amplitude
from the $i$-th input port to the $j$-th output port, that is 
$b_j = \sum_{k=1}^3 T_{jk} a_k$. \par
Such devices have already been implemented in single-mode optical
fiber technology and commercial triple coupler have been available for
some time~\cite{she}. Any triple coupler can be also implemented by discrete
optical components using symmetric beam splitters and phase shifters
only \cite{zap}. As it has already mentioned in Section \ref{s:eig}, 
this is due to remarkable mathematical fact that that any
unitary $M$-dimensional matrix can be factorized into a sequence of
2-dimensional transformations plus phase-shift \cite {zei}. 
Moreover, this decomposition
is not, in general, unique. In Fig.~\ref{f:TTT} a possible implementation of a
triple coupler is schematically reported.
Experimental realizations of triple couplers  has been reported for both cases,
the passive elements case and the optical fiber one \cite{zap,zin,ol1,ol2}.
\par
Let us now consider the measurement scheme of Fig.~\ref{f:THD}.
The three input modes are mixed by a triple coupler and the resulting output
modes are subsequently detected by three identical photodetectors. The
measured photocurrents are proportional to $\hat I_n$, $n=1,2,3$ given by
\begin{eqnarray}
\fl \hat I_n &=& b^{\dag}_n b_n = \frac{1}{3} \sum_{k,l=1}^3
\exp\left\{i\theta_n (l-k)\right\} a_k^{\dag} a_l\;, \qquad
\theta_n=\frac{2\pi}{3}(n-1)
\label{pht}\; .
\end{eqnarray}
After photodetection a Fourier transform (FT) on the photocurrents is 
performed
\begin{eqnarray}
\fl \hat {\cal I}_s \equiv {\rm FT}(\hat I_1,\hat I_2,\hat I_3) 
= \frac{1}{\sqrt{3}} \sum_{n=1}^3 \hat I_n
\exp\left\{-i\theta_n (s-1)\right\}\; , \qquad s=1,2,3 \, .
\label{FT1}
\end{eqnarray}
This procedure is a straightforward generalization of the customary
two-mode balanced homodyning technique. In that case, in fact, the sum
and the difference of the two output photocurrents are considered, which
actually represent the Fourier transform in a two-dimensional space.
By means of the identity
\begin{eqnarray}
\delta_3 (s-1)=\frac{1}{3}\sum_{n=1}^3\exp\left\{i\frac{2\pi}{3}n(s-1)\right\}
\label{FT2}\;,
\end{eqnarray}
for the periodic (modulus 3) Kronecker delta  $\delta_3 $, we  
obtain our final expressions for the Fourier transformed photocurrents
\begin{eqnarray}
   \hat {\cal I}_1 = \frac{1}{\sqrt{3}} \left\{
a^{\dag}_1 a_1 + a^{\dag}_2 a_2 + a^{\dag}_3 a_3\right\} \; ,
\\ \hat {\cal I}_2 =\frac{1}{\sqrt{3}} \left\{
a^{\dag}_1 a_2 + a^{\dag}_2 a_3 + a^{\dag}_3 a_1\right\} \; ,
\\ \hat {\cal I}_3 =
\frac{1}{\sqrt{3}} \left\{a^{\dag}_1 a_3 +a^{\dag}_2 a_1+ a^{\dag}_3 a_2
\right\}\label{FT3}\;.
\end{eqnarray}
$\hat {\cal I}_1 $ gives no relevant information as it is insensitive to the phase
of the signal field, whereas $\hat {\cal I}_2 $ and $\hat {\cal I}_3$
are hermitian conjugates of each other and contain the relevant information
in their real and imaginary part. 
In the following let us assume $a_1$ is the signal mode and $a_2$
is fed by a highly excited coherent state $| z \rangle$ representing 
the local oscillator. For large $|z|$ the output photocurrents are intense
enough to be easily detected. They can be combined to give the reduced
photocurrents
\begin{eqnarray}
\widehat {\cal Z}_1 &=& \sqrt{3}\frac{\hat {\cal I}_2 +
\hat {\cal I}_3 }{2 |z|} = \hat a_{1} ( 0) + \hat a_{3}( 0 )
+ O[\frac{1}{|z|}] \nonumber \\ \widehat {\cal Z}_2 &=& \sqrt{3}
\frac{\hat {\cal I}_2 - \hat {\cal I}_3 }{2 i|z|}=-\hat a_{1} (\pi/2)
+ \hat a_{3} (\pi/2) + O[\frac{1}{|z|}] \label{THP}\;,
\end{eqnarray}
which we refer to as the {\em triple homodyne photocurrents}.
Again the complex photocurrent $\widehat {\cal Z}=
\widehat {\cal Z}_1+ i \widehat {\cal Z}_2$ has the form
$\widehat {\cal Z}= a_1+ a_3^{\dag}$, being $a_1$ the signal mode and
$a_3$ the idler of the device. \par
When accounting for the non unit quantum efficiency $\eta$ of the
photodetectors the output modes are written as
$$b_j = \sqrt{\eta}\:\left(\sum_{k=1}^3 T_{jk} a_k\right)
+ \sqrt{1-\eta}\: u_j  \qquad j=1,2,3\:,$$
so that the reduced photocurrents are now given by
\begin{eqnarray}
\fl \widehat {\cal Z}_1= \sqrt{3}\frac{\hat {\cal I}_2 +
\hat {\cal I}_3 }{2 \eta|z|} = \hat a_{1} ( 0) + \hat a_{3}( 0 )
+ \sqrt{\frac{1-\eta}{\eta}} \left[\hat u_{1} (0 ) -
\hat u_{3} (0)\right] + O[\frac{1}{|z|}]
\nonumber \\ \fl \widehat {\cal Z}_2=
\sqrt{3} \frac{\hat {\cal I}_2 - \hat {\cal I}_3 }{2 i\eta |z|}=
-\hat a_{1} (\pi/2) + \hat a_{3} (\pi/2)
+ \sqrt{\frac{1-\eta}{\eta}} \left[ \hat u_{1} (\pi/2) - \hat u_{3} (\pi/2)
\right] + O[\frac{1}{|z|}]
\label{6zphot}\:.
\end{eqnarray}
When, as it is the case, the modes $u_j$ are placed in the vacuum the 
six-port photocurrents in Eq. (\ref{6zphot}) leads to the same statistics of 
the eight-port photocurrents in Eq. (\ref{zphot}). Indeed, they describe 
different devices leading to the same amount of information on the signal 
mode $a_1$. 
Some comments are, however, in order. By comparison of Fig. \ref{f:rough} and 
Fig. \ref{f:THD} it appears obvious that six-port homodyne is an optimized scheme relative to the eight-port one. One mode less is needed to reach the same
amount of information, thus decreasing the possible sources of noise.
The reason for this lies in the final stage of the two schemes. 
The Fourier transform of the six-port photocurrents, in fact, is a more 
effective procedure relative to the double-homodyne final stage of the 
eight-port scheme.
This is related to the noise suppression mechanism of homodyne detection. 
The latter, in fact, is generalized to the multi-mode case by the Fourier 
transform rather than duplication of the original two-mode scheme.
\section{Output statistics from a two-photocurrent device}
\label{s:pom}
In this section we analyze with some detail the output statistics
of an abstract two-photocurrent device. The latter is
characterized by the joint measurement of the real $\widehat{\cal Z}_1$ and
the imaginary $\widehat{\cal Z}_2$ part of the complex photocurrent
\begin{eqnarray}\widehat{\cal Z} = a + b^{\dag} \:,\label{ZED}\end{eqnarray}
when equipped with perfect photodetectors. On the other hand, in the
realistic case of inefficient photodetection the photocurrent is given by
\begin{eqnarray}\widehat{\cal Z} = a + b^{\dag} + \sqrt{\frac{1-\eta}{\eta}}
\left( u_1 + u_2^{\dag}\right)\:.\label{Zeta}\end{eqnarray}
In Eqs. (\ref{ZED}) and (\ref{Zeta}) $a$ and $b$ denote single mode
radiation field which can be excited in any quantum state. The operators
$u_1$ and $u_2$ also denote single mode radiation field, however
strictly placed in the vacuum states. They are used to simulate losses
due inefficient photodetection, according to the equivalent detection
scheme of Section \ref{s:eta}.
In the following we will refer to the mode $a$ as the signal mode which
is under examination, whereas the mode $b$ is in a known and fixed state, 
playing the role of the probe mode of the device. This is only for sake of
convenience. It is obvious that the roles of the two modes can be
interchanged and any argument can be reversed considering the mode $b$ as
a signal mode. According to this scheme we denote the outcome
probability density distribution by $K_b(\alpha,\bar\alpha)$. The latter
describes, in the complex plane, the state of the mode $a$ as probed
by the mode $b$. Indeed, different choices for the probe mode
lead to very different features in the probability distribution. \par
Each experimental random outcome $(z_1,z_2)$ from the joint measurement of
$\widehat{\cal Z}_1$ and $\widehat{\cal Z}_2$ can be considered as a
point $z$ in the complex plane. On the other hand, the output photocurrent
$\widehat{\cal Z}$ is expressed as a sum of different contributions
coming from different modes. Therefore, it appears intuitively rather
obvious that the resulting probability distribution will be given by
a convolution. To be more specific let us start by considering the ideal
case of unit quantum efficiency $\eta=1$. We write the probability
distribution $K_b(\alpha,\bar\alpha)$ as the Fourier transform
\begin{eqnarray}
K_b(\alpha,\bar\alpha) = \int_{\bf C} \frac{d^2\gamma}{\pi^2}\:
e^{\bar{\gamma}\alpha-\gamma\bar\alpha}\:\Xi (\gamma,\bar\gamma)
\label{FT}\:,
\end{eqnarray}
of the characteristic function
\begin{eqnarray}
\Xi (\gamma,\bar\gamma)=\hbox{Tr}\left\{\hat\varrho\:\exp\left[\bar\gamma
\widehat{\cal Z}-\gamma\widehat{\cal Z}^{\dag}\right]\right\}\label{Xi}\:,
\end{eqnarray}
being $\hat\varrho$ the global density matrix describing both modes $a$ and
$b$. We consider probe mode to be independent on the signal mode, so
that the input mode is factorized as
\begin{eqnarray}\hat\varrho=\hat\varrho_a
\otimes\hat\varrho_b\label{rho}\:.\end{eqnarray}
Upon substituting Eqs. (\ref{ZED}) and (\ref{rho}) in Eq. (\ref{Xi}) we are
able to write the characteristic function $\Xi(\gamma,\bar\gamma)$
as a product
\begin{eqnarray}
\Xi (\gamma,\bar\gamma)= \hbox{Tr}\left\{\hat\varrho_a \otimes
\hat\varrho_b \;\; \hat D_a(\gamma) \otimes \hat D_b (-\gamma) \right\}
= \chi_a(\gamma,\bar{\gamma})\;\chi_b(-\gamma,-\bar{\gamma})
\label{pXi}\:,
\end{eqnarray}
being $$\hat D(\gamma) = \left\{\hat\rho \exp\left[ \gamma a^{\dag} -
\bar{\gamma}a \right] \right\}$$ the displacement operator and 
$$\chi(\gamma, \bar\gamma)= \hbox{Tr}\left\{\hat\varrho\:\hat D(\gamma)\right\}$$ the single
mode characteristics function, the latter entering in the definition of
the Wigner function of a single mode radiation field \cite{wig,gl1,gl2}
\begin{eqnarray}
W(\alpha,\bar\alpha )=\int\frac{d^2\lambda}{\pi}\:\chi(\lambda,\bar{\lambda})
\:\exp\left\{\bar\lambda \alpha -\lambda \bar\alpha\right\} \label{Wdf}\;.
\end{eqnarray}
We now insert Eq. (\ref{pXi}) in Eq. (\ref{FT}). By means of Eq.
(\ref{Wdf}) and using the convolution theorem we arrive at the final
result
\begin{eqnarray}\fl K_b(\alpha,\bar\alpha )= W_a(\alpha,\bar\alpha )\: \star
\:W_b(-\alpha,-\bar\alpha ) = \int_{\bf C} \frac{d^2 \beta}{\pi^2} \;
W_b(\alpha+\beta,\bar\alpha +\bar\beta )\: W_a(\beta,\bar\beta )
\label{Con}\;,\end{eqnarray}
the symbol $\star$ denoting convolution.
Eq. (\ref{Con}) provides a justification for referring to the mode $b$
as the probe of the device. In fact, it acts as a filter on the signal'
Wigner function. The latter is not a genuine probability distribution, 
as it can be negative when describing quantum interference. Thus, it cannot be 
directly sampled by experiments. However, the convolution
(\ref{Con}) make it more regular, leading to $K_b(\alpha,\bar\alpha)$
which is a measurable distribution. \par 
Phase space density as in Eq. (\ref{Con}) have been already introduced
by Wodkiewicz \cite{wo1,wo2,wo3} to account for the effect of the measuring
apparatus in a joint measurement of conjugated variables. Originally,
they have been termed phase-space {\em propensities}. 
More recently, they also have been 
used in entropic description of quantum mechanical states \cite{bu1,bu2}.
Two-photocurrent devices appear as the natural setup to start from, 
in order to experimentally access such kind of phase-space distribution. 
\par Wigner function, though it contains a complete description of 
the quantum state, cannot be directly measured. The phase distributions 
$K_b(\alpha,\bar\alpha)$ are smoothed version of it, corresponding to the
occurrence of additional noise of purely quantum origin. Indeed, 
two-photocurrent devices provide the generalized joint measurement of 
position and momentum, thus unavoidably introducing additional noise 
by first principles \cite{y82,hef,art}. The crucial point to notice 
here is that they offer the remarkable possibility to manipulate this 
quantum noise. As it emerges from Eq. (\ref{Con}) it can be redirected
in the desidered region of the complex plane by suitable choice of the probe 
mode, according to the kind of information of which is of interest. \par 
In order to incorporate the effects of inefficient detection we first note
that Eq. (\ref{Zeta}) differs from Eq. (\ref{ZED}) by two additional 
{\em additive} terms. Therefore, further convolutions to the original 
Wigner function are expected. Indeed, the characteristic function of the 
whole device is now expressed as 
\begin{eqnarray}
\Xi_{\eta} (\gamma,\bar\gamma) &=& \chi_a(\gamma,\bar{\gamma})\;
\chi_b(-\gamma,-\bar{\gamma})\times \label{pXieta} \\ \nonumber \\
&\times&\chi_{u_1}\left(\sqrt{\frac{1-\eta}{\eta}}
\gamma,\sqrt{\frac{1-\eta}{\eta}}\bar{\gamma}\right)\;
\chi_{u_2}\left(-\sqrt{\frac{1-\eta}{\eta}}\gamma,-\sqrt{\frac{1-\eta}{\eta}}
\bar{\gamma}\right)\nonumber
\end{eqnarray}
where characteristic function for the noise modes can be easily evaluated as
\begin{eqnarray}
\fl
\chi_{u_j}\left(-\sqrt{\frac{1-\eta}{\eta}}\gamma,-\sqrt{\frac{1-\eta}{\eta}}
\bar{\gamma}\right)=\exp\left[-\frac{1-\eta}{2\eta}|\gamma |^2\right]
\qquad j=1,2
\label{noiseXi}\:.
\end{eqnarray}
Upon inserting Eqs. (\ref{pXieta}) and (\ref{noiseXi}) in Eq. (\ref{FT})
we obtain our final result, namely the output probability distribution
$K_{b\eta}(\alpha ,\bar{\alpha}) $ of a two-photocurrent device equipped 
with photodetectors of quantum efficiency $\eta$
\begin{eqnarray}
\fl K_{b\eta}(\alpha,\bar\alpha )= K_{b1}(\alpha,\bar\alpha )
\: \star \: G_{\eta}(\alpha,\bar\alpha ) = 
\int_{\bf C} \frac{d^2 \beta}{\pi^2} \; K_b(\beta,\bar\beta )\: 
\exp\left[-\frac{\eta}{1-\eta}|\alpha -\beta|^2\right]
\label{noiseCon}\;,\end{eqnarray}
being $K_{b1}(\alpha,\bar\alpha )$ the probability distribution 
obtained for ideal photodetection and 
$$G_{\eta}(\alpha,\bar\alpha )= \exp\left[-\frac{\eta}{1-\eta}
|\alpha |^2\right]$$ the filter function summarizing the effects of 
inefficient photodetection. 
\section{Conclusions}
\label{s:out}
In this paper we have introduced the class of two-photocurrent devices.
This kind of detectors are characterized by the fact that they jointly 
measure the real and the imaginary part of the complex photocurrent
${\cal Z}=a + b^{\dag}$, where $a$ and $b$ describe two single mode of the  
radiation field. Eight-port homodyne, heterodyne and six-port homodyne
detectors belong to this general class, thus their fully equivalence 
in probing radiation field has been proved. It has also been proved that
this equivalence still holds when the inefficiency of the photodetection 
process is taken into account. This is an interesting and unexpected result, 
as photodetection takes place at very different stages in the three 
detection schemes. \par
The three schemes analyzed in this paper are equivalent from the point 
of view of provided information on the measured signal.
Nevertheless, they have different physical implementations which have 
to be compared. We pointed out the advantages of the 
six-port homodyne scheme in comparison with the customary eight-port one.
Actually, it provides the minimal scheme to access generalized phase 
space distribution \cite{new}. \par
In a two-photocurrent device a generalized joint measurement of position
and momentum is performed on the signal mode. This results in a smoothing
of the signal Wigner function to a measurable distribution, which represents 
the output probability distribution of the measurement. 
Some additional noise is unavoidably introduced, according to the Heisenberg 
principle for joint measurement. However, the filtering process has been 
shown to be a convolution with the Wigner function of the probe mode. 
Therefore, it is possible to manipulate and redirect the noise.
A suitable choice of the probe mode enhances different features of the signal' 
phase space distribution, according to the kind of desidered information. 
\section*{Acknowledgment}
I would thank Prof. Harry Paul for his kind hospitality in the
group 'Nichtklassiche Strahlung' of Max-Planck society and Prof. 
Mauro D'Ariano for introducing me to the subject of phase-space 
measurements. I would also thank Valentina De Renzi, Ole Steuernagel, 
Gabriel Drobny, Alexei Chizhov and Alfred W\"unsche for profitable 
discussions. This work has been partially supported by the University 
of Milan by a scholarship for postgraduate studies in foreign countries.

\vfill\newpage
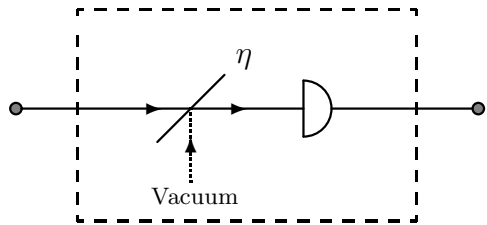
\begin{figure}[ht]
\setlength{\unitlength}{0.75cm}
\begin{center}\begin{picture}(10,4)(-4,-2)\thicklines
\drawline(-3,0)(2,0)\put(-3.1,0){\shade\circle{.2}}\drawline(-.6,-.6)(.6,.6)
\put(.8,.8){$\eta$}\put(2,0){\arc{1}{4.71}{1.57}}\drawline(2,-.5)(2,.5)
\put(5.1,0){\shade\circle{.2}}\drawline(2.5,0)(5,0)\put(0,-1.3){\dashbox{.05}(0,1.3)}
\put(-.7,-1.7){\small\sc Vacuum}\put(0,-.5){\vector(0,1){0}}\put(-.5,0){\vector(1,0){0}}
\put(1,0){\vector(1,0){0}}\put(-2.0,-2.0){\dashbox{.2}(6,3.75)}
\end{picture}\end{center}
\caption{Equivalent scheme for inefficient photodetection.}\label{f:eta}
\end{figure}
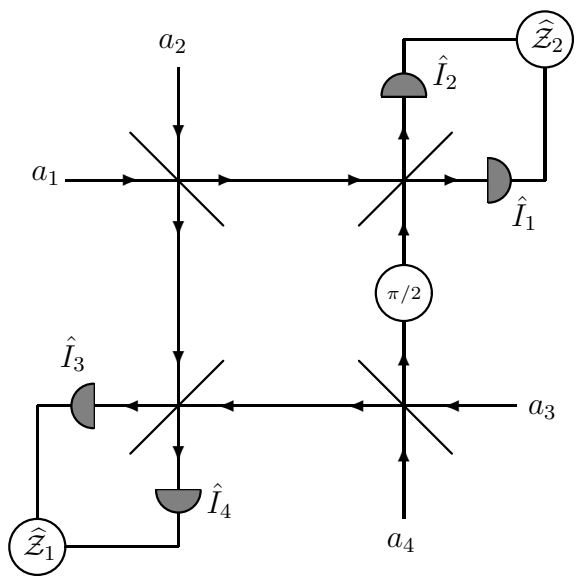
\begin{figure}[h]\begin{center}\setlength{\unitlength}{0.75cm}
\begin{picture}(10,10)(-5,-5)\thicklines
\put(-2,-4.5){\line(0,1){0.6}}\put(-2,-3.5){\line(0,1){7.5}}
\put(-4,2){\line(1,0){2}}\put(-4.5,-2){\line(1,0){0.6}}
\put(-1.2,1.2){\line(-1,1){1.65}}\put(1.2,-1.2){\line(1,-1){1.65}}
\put(-1.2,-1.2){\line(-1,-1){1.65}}\put(1.2,1.2){\line(1,1){1.65}}
\put(-3.5,-2){\line(1,0){7.5}}\put(-4,-4.5){\line(1,0){2}}
\put(-4.5,-4){\line(0,1){2}}\put(2,-4){\line(0,1){3.5}}
\put(2,.5){\line(0,1){3}}\put(2,3.9){\line(0,1){0.6}}
\put(-4,2){\line(1,0){7.5}}\put(3.9,2){\line(1,0){0.6}}
\put(2,4.5){\line(1,0){2}}\put(4.5,2){\line(0,1){2}}
\put(2,3.5){\line(1,0){.4}}\put(2,3.5){\line(-1,0){.4}}
\put(-2,-3.5){\line(1,0){.4}}\put(-2,-3.5){\line(-1,0){.4}}
\put(-3.5,-2){\line(0,1){.4}}\put(-3.5,-2){\line(0,-1){.4}}
\put(3.5,2){\line(0,1){.4}}\put(3.5,2){\line(0,-1){.4}}
\put(-4.5,-4.5){\circle{1}}\put(2,0){\circle{1}}\put(4.5,4.5){\circle{1}}
\put(2,3.5){\shade\arc{.8}{3.141}{0}}\put(-2,-3.5){\shade\arc{.8}{0}{3.141}}
\put(-3.5,-2){\shade\arc{.8}{1.57}{4.71}}\put(3.5,2){\shade\arc{.8}{4.71}{1.57}}
\put(-2,-1.3){\vector(0,-1){0}}\put(-2,-3){\vector(0,-1){0}}
\put(1,-2){\vector(-1,0){0}}\put(2.7,-2){\vector(-1,0){0}}
\put(-1.3,-2){\vector(-1,0){0}}\put(-3,-2){\vector(-1,0){0}}
\put(-2,1){\vector(0,-1){0}}\put(-2,2.7){\vector(0,-1){0}}
\put(2,1.3){\vector(0,1){0}}\put(2,3){\vector(0,1){0}}
\put(1.3,2){\vector(1,0){0}}\put(3,2){\vector(1,0){0}}
\put(-1,2){\vector(1,0){0}}\put(-2.7,2){\vector(1,0){0}}
\put(2,-1){\vector(0,1){0}}\put(2,-2.7){\vector(0,1){0}}
\put(4.25,4.35){\small{$\widehat{\cal Z}_2$}}
\put(-4.75,-4.65){\small{$\widehat{\cal Z}_1$}}
\put(1.7,-.1){\tiny{$\pi/2$}}
\put(4.2,-2.15){$a_3$}\put(-4.6,2){$a_1$}
\put(-1.5,-3.9){$\hat I_4$}\put(-4.1,-1.3){$\hat I_3$}
\put(2.5,3.7){$\hat I_2$}\put(3.9,1.2){$\hat I_1$}
\put(1.7,-4.5){$a_4$}\put(-2.35,4.35){$a_2$}
 \end{picture}
\end{center}
\caption{Schematic diagram of an eight-port homodyne detector.}
\label{f:eig}
\end{figure}
\begin{figure}[ht]
\begin{center}\unitlength=1mm
\begin{picture}(110.00,60.00)
\put(25.00,11.00){\rule{27.00\unitlength}{49.00\unitlength}}
\drawline(15.00,20.00)(25.00,20.00)\drawline(15.00,30.00)(25.00,30.00)
\drawline(15.00,40.00)(25.00,40.00)\drawline(15.00,50.00)(25.00,50.00)
\put(14.00,50.00){\circle*{2.00}}\put(14.00,40.00){\circle*{2.00}}
\put(14.00,30.00){\circle*{2.00}}\put(14.00,20.00){\circle*{2.00}}
\drawline(52.00,50.00)(65.00,50.00)\drawline(65.00,40.00)(52.00,40.00)
\drawline(52.00,30.00)(65.00,30.00)\drawline(65.00,20.00)(52.00,20.00)
\drawline(65.00,47.00)(65.00,53.00)\drawline(65.00,43.00)(65.00,37.00)
\drawline(65.00,33.00)(65.00,27.00)\drawline(65.00,23.00)(65.00,17.00)
\put(65,20){\shade\arc{6}{4.71}{1.57}}\put(65,30){\shade\arc{6}{4.71}{1.57}}
\put(65,40){\shade\arc{6}{4.71}{1.57}}\put(65,50){\shade\arc{6}{4.71}{1.57}}
\drawline(68.00,50.00)(75.00,50.00)\drawline(75.00,40.00)(68.00,40.00)
\drawline(68.00,30.00)(75.00,30.00)\drawline(75.00,20.00)(68.00,20.00)
\drawline(89.00,45.00)(100.00,45.00)\drawline(81.00,23.00)(75.00,20.00)
\put(85.00,45.00){\circle{8.00}}\put(85.00,25.00){\circle{8.00}}
\put(100.00,45.00){\circle*{2.00}}\put(100.00,25.00){\circle*{2.00}}
\drawline(100.00,25.00)(89.00,25.00)\drawline(75.00,50.00)(81.00,47.00)
\drawline(75.00,40.00)(81.00,43.00)\drawline(75.00,30.00)(81.00,27.00)
\put(9.,53.){\makebox(0,0)[cc]{$a_1$}}\put(9.,43.){\makebox(0,0)[cc]{$a_2$}}
\put(9.,33.){\makebox(0,0)[cc]{$a_3$}}\put(9.,23.){\makebox(0,0)[cc]{$a_4$}}
\put(57,53.){\makebox(0,0)[cc]{$b_1$}}\put(57,43.){\makebox(0,0)[cc]{$b_2$}}
\put(57,33.){\makebox(0,0)[cc]{$b_3$}}\put(57,23.){\makebox(0,0)[cc]{$b_4$}}
\put(72,53.){\makebox(0,0)[cc]{$\hat I_1$}}
\put(72,43.){\makebox(0,0)[cc]{$\hat I_2$}}
\put(72,33.){\makebox(0,0)[cc]{$\hat I_3$}}
\put(72,23.){\makebox(0,0)[cc]{$\hat I_4$}}
\put(105.00,45.00){\makebox(0,0)[lc]{${\cal Z}_1$}}
\put(105.00,25.00){\makebox(0,0)[lc]{${\cal Z}_2$}}
\put(85,45){\makebox(0,0)[cc]{\bf --}}\put(85,25){\makebox(0,0)[cc]{\bf --}}
\end{picture}\end{center}
\vspace{-25pt}
\caption{Eight-port homodyne detector as a multiport homodyne.}
\label{f:rough}
\end{figure}
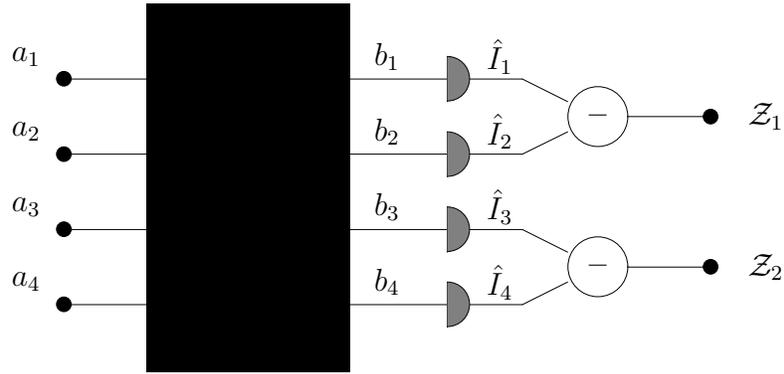
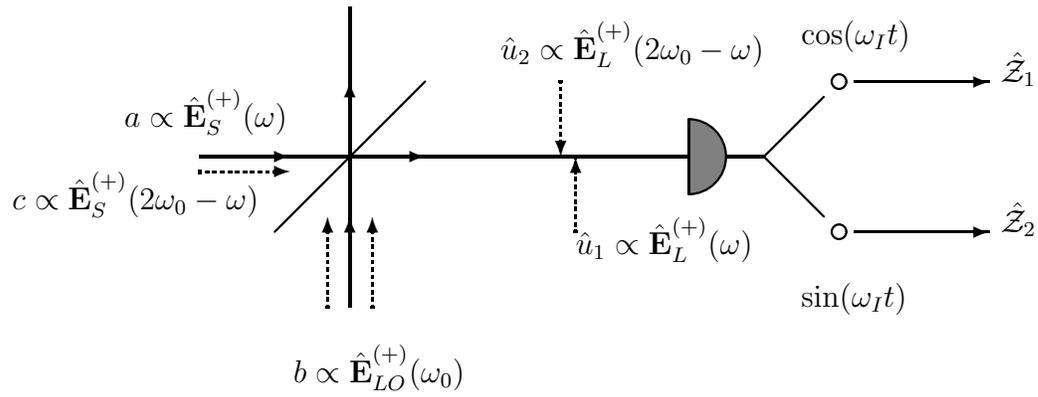
\begin{figure}[ht]\setlength{\unitlength}{1cm}
\begin{center}\begin{picture}(14,7)(-4.5,-3.5)\thicklines
\put(-2,0){\line(1,0){6.5}}\put(-2,-0.2){\dashbox{.05}(1,0)}
\put(-.8,-.2){\vector(1,0){0}}\put(0,-2){\line(0,1){4}}
\put(-.3,-2){\dashbox{.05}(0,1)}\put(.3,-2){\dashbox{.05}(0,1)}
\put(-.3,-.8){\vector(0,1){0}}\put(.3,-.8){\vector(0,1){0}}
\put(-1,-1){\line(1,1){2}}\put(-.8,0){\vector(1,0){0}}
\put(0,-.8){\vector(0,1){0}}\put(1,0){\vector(1,0){0}}
\put(0,1){\vector(0,1){0}}\put(4.5,0){\shade{\arc{1.}{-1.5708}{1.5708}}}
\put(4.5,-.5){\line(0,1){1}}\put(5,0){\line(1,0){.5}}
\put(5.5,0){\line(1,1){.8}}\put(5.5,0){\line(1,-1){.8}}
\put(6.5,1){\circle{.2}}\put(6.5,-1){\circle{.2}}
\put(6.7,1){\line(1,0){1.8}}\put(6.7,-1){\line(1,0){1.8}}
\put(8.5,1){\vector(1,0){0}}\put(8.5,-1){\vector(1,0){0}}
\put(-3.,.4){$a \propto \hat{\E}^{(+)}_S (\omega)$}
\put(-4.49,-.7){$ c \propto \hat{\E}^{(+)}_S (2\omega_0-\omega )$}
\put(-.75,-3){$b \propto \hat{\E}^{(+)}_{LO} (\omega_0 )$}
\put(3,-1){\dashbox{.05}(0,1)}\put(3,0){\vector(0,1){0}}
\put(3,-1.3){$\hat u_1 \propto \hat{\E}^{(+)}_L (\omega ) $}
\put(2.8,0){\dashbox{.05}(0,1)}\put(2.8,0){\vector(0,-1){0}}
\put(2,1.3){$\hat u_2 \propto \hat{\E}^{(+)}_L (2\omega_0-\omega )$}
\put(6,1.5){$\cos (\omega_{I} t)$}\put(6,-2){$\sin (\omega_{I} t)$}
\put(8.65,1){$\hat {\cal Z}_1$}\put(8.65,-1){$\hat {\cal Z}_2$}
\end{picture}\end{center}
\caption{Schematic diagram of a heterodyne detection.
Relevant modes are explicitly pointed out.}\label{f:het}
\end{figure}
\begin{figure}[ht]
\begin{center}\unitlength=1mm
\begin{picture}(143.33,53.00)(25.00,25.00)
\drawline(50.33,50.00)(55.33,50.00)\drawline(50.33,40.00)(55.33,40.00)
\drawline(50.33,30.00)(55.33,30.00)\drawline(55.33,30.00)(65.33,40.00)
\drawline(55.33,40.00)(65.33,30.00)\drawline(65.33,40.00)(70.33,40.00)
\drawline(65.33,30.00)(70.33,30.00)\drawline(55.33,50.00)(70.33,50.00)
\drawline(78.33,30.00)(80.33,30.00)\drawline(80.33,30.00)(80.33,30.00)
\drawline(80.33,30.00)(85.33,30.00)\drawline(80.33,40.00)(85.33,40.00)
\drawline(80.33,50.00)(85.33,50.00)\drawline(80.33,50.00)(70.33,40.00)
\drawline(70.33,50.00)(80.33,40.00)\drawline(70.33,30.00)(72.33,30.00)
\drawline(85.33,50.00)(87.33,50.00)\drawline(85.33,40.00)(95.33,40.00)
\drawline(85.33,30.00)(95.33,30.00)\drawline(93.33,50.00)(95.33,50.00)
\drawline(95.33,50.00)(100.33,50.00)\drawline(95.33,40.00)(100.33,40.00)
\drawline(95.33,30.00)(100.33,30.00)\drawline(100.33,40.00)(110.33,50.00)
\drawline(100.33,50.00)(110.33,40.00)\drawline(110.33,50.00)(120.33,50.00)
\drawline(110.33,40.00)(120.33,40.00)\drawline(100.33,30.00)(120.33,30.00)
\drawline(120.33,30.00)(130.33,40.00)\drawline(120.33,40.00)(130.33,30.00)
\drawline(130.33,30.00)(140.33,30.00)\drawline(130.33,40.00)(140.33,40.00)
\drawline(120.33,50.00)(140.33,50.00)\put(90.33,50.00){\circle{6.00}}
\put(75.33,30.00){\circle{6.00}}\put(50.33,50.00){\circle*{2.00}}
\put(50.33,40.00){\circle*{2.00}}\put(50.33,30.00){\circle*{2.00}}
\put(140.33,50.00){\circle*{2.00}}\put(140.33,40.00){\circle*{2.00}}
\put(140.33,30.00){\circle*{2.00}}
\put(57.33,34.00){\rule{6.00\unitlength}{2.00\unitlength}}
\put(72.33,44.00){\rule{6.00\unitlength}{1.00\unitlength}}
\put(72.33,44.00){\rule{6.00\unitlength}{2.00\unitlength}}
\put(102.33,44.00){\rule{6.00\unitlength}{2.00\unitlength}}
\put(122.33,34.00){\rule{6.00\unitlength}{2.00\unitlength}}
\put(75.66,30.00){\makebox(0,0)[cc]{$\varphi_1$}}
\put(90.66,50.00){\makebox(0,0)[cc]{$\varphi_2$}}
\put(145,50.00){\makebox(0,0)[cc]{$\hbox{b}_1$}}
\put(145,40.00){\makebox(0,0)[cc]{$\hbox{b}_2$}}
\put(145,30.00){\makebox(0,0)[cc]{$\hbox{b}_3$}}
\put(46.33,50.00){\makebox(0,0)[cc]{$\hbox{a}_1$}}
\put(46.33,40.00){\makebox(0,0)[cc]{$\hbox{a}_2$}}
\put(46.33,30.00){\makebox(0,0)[cc]{$\hbox{a}_3$}}
\put(67.00,36.00){\makebox(0,0)[cc]{\footnotesize\sc BS}}
\put(82.00,46.00){\makebox(0,0)[cc]{\footnotesize\sc BS}}
\put(112.00,46.00){\makebox(0,0)[cc]{\footnotesize\sc BS}}
\put(132.00,36.00){\makebox(0,0)[cc]{\footnotesize\sc BS}}
\end{picture}\end{center}
\caption{Realization of a triple coupler in terms of 50:50 beam splitters
(BS) and phase shifters '$\varphi$'. In order to obtain a symmetric coupler
the following values has to be chosen: $\varphi_1=\arccos (1/3)$ and
$\varphi_2 = \varphi_1 /2$.}\label{f:TTT}
\end{figure}
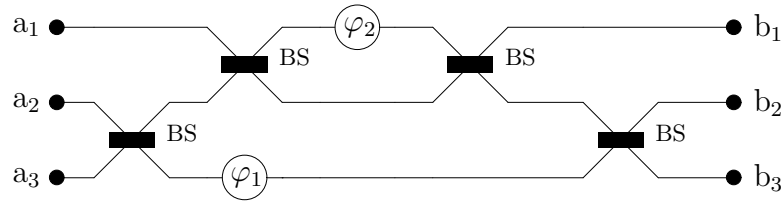
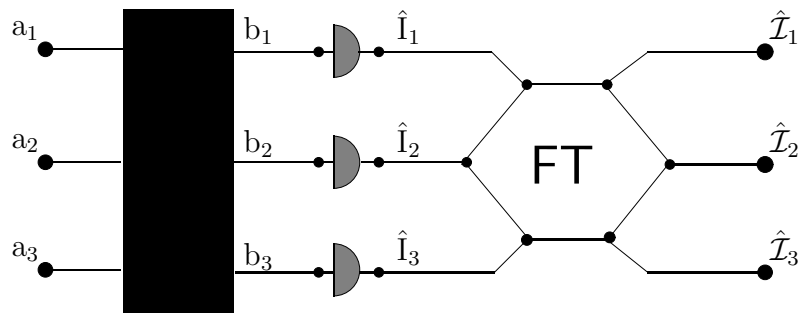
\begin{figure}[ht]
\begin{center}\unitlength=1mm
\begin{picture}(107.67,56.66)
\put(19.67,16.33){\rule{14.67\unitlength}{40.33\unitlength}}
\put(9.34,36.33){\line(1,0){10.00}}\put(64.67,36.00){\line(5,6){9.00}}
\put(73.67,46.66){\line(1,0){10.00}}\put(83.67,46.66){\line(5,-6){8.67}}
\put(92.34,36.33){\line(-4,-5){8.33}}\put(84.00,26.00){\line(-1,0){10.33}}
\put(73.67,26.00){\line(-5,6){8.33}}\put(83.67,46.66){\line(4,3){5.67}}
\put(89.34,51.00){\line(1,0){15.33}}\put(92.00,36.00){\line(1,0){12.33}}
\put(84.00,26.00){\line(5,-4){5.33}}\put(89.34,21.66){\line(1,0){15.33}}
\put(73.34,26.00){\line(-1,-1){4.33}}\put(9.34,51.33){\circle*{1.49}}
\put(9.34,36.33){\circle*{1.33}}\put(9.34,21.99){\circle*{2.00}}
\put(9.34,36.33){\circle*{2.00}}\put(9.34,51.33){\circle*{2.00}}
\put(105.00,51.00){\circle*{2.11}}\put(105.00,36.00){\circle*{2.11}}
\put(105.00,21.66){\circle*{2.00}}\put(73.34,46.66){\circle*{1.33}}
\put(84.00,46.66){\circle*{1.33}}\put(92.34,36.00){\circle*{1.33}}
\put(84.34,26.33){\circle*{1.49}}\put(65.34,36.33){\circle*{1.33}}
\put(73.34,26.00){\circle*{1.49}}
\put(78.00,36.00){\makebox(0,0)[cc]{\LARGE \sf FT}}
\put(6.67,53.83){\makebox(0,0)[cc]{$\hbox{a}_1$}}
\put(6.67,38.83){\makebox(0,0)[cc]{$\hbox{a}_2$}}
\put(6.67,24.44){\makebox(0,0)[cc]{$\hbox{a}_3$}}
\put(37.67,53.83){\makebox(0,0)[cc]{$\hbox{b}_1$}}
\put(37.67,38.83){\makebox(0,0)[cc]{$\hbox{b}_2$}}
\put(37.67,23.83){\makebox(0,0)[cc]{$\hbox{b}_3$}}
\put(57.67,54.33){\makebox(0,0)[cc]{$\hat{\hbox{I}}_1$}}
\put(57.67,39.33){\makebox(0,0)[cc]{$\hat{\hbox{I}}_2$}}
\put(57.67,24.99){\makebox(0,0)[cc]{$\hat{\hbox{I}}_3$}}
\put(107.67,54.33){\makebox(0,0)[cc]{$\hat{\cal I}_1$}}
\put(107.67,39.33){\makebox(0,0)[cc]{$\hat{\cal I}_2$}}
\put(107.67,24.99){\makebox(0,0)[cc]{$\hat{\cal I}_3$}}
\put(68.67,51.00){\line(1,-1){4.67}}\put(9.34,51.33){\line(1,0){10.33}}
\put(9.34,22.00){\line(1,0){10.00}}\put(44.67,36.33){\line(-1,0){12.00}}
\put(41.34,36.33){\line(1,0){6.30}}
\put(47.67,36.33){\shade{\arc{7.}{-1.5708}{1.5708}}}
\put(47.67,22.00){\shade{\arc{7.}{-1.5708}{1.5708}}}
\put(47.67,51.08){\shade{\arc{7.}{-1.5708}{1.5708}}}
\put(47.67,47.58){\line(0,1){7.00}}\put(47.67,32.83){\line(0,1){7.00}}
\put(47.67,18.50){\line(0,1){7.00}}\put(51.34,36.33){\line(1,0){14.00}}
\put(68.67,51.00){\line(-1,0){17.67}}\put(69.00,21.66){\line(-1,0){17.89}}
\put(51.67,21.66){\line(0,1){0.00}}\put(47.67,51.00){\line(-1,0){13.67}}
\put(47.67,21.66){\line(-1,0){13.00}}\put(45.67,51.00){\circle*{1.33}}
\put(45.67,21.66){\circle*{1.33}}\put(45.67,36.33){\circle*{1.33}}
\put(53.67,51.00){\circle*{1.33}}\put(53.67,36.33){\circle*{1.33}}
\put(53.67,21.66){\circle*{1.33}}
\end{picture}\end{center} \vspace{-40pt}
\caption{Outline of triple coupler homodyne detectors: The hexagonal box
symbolizes the electronically performed Fourier transform (FT).}\label{f:THD}
\end{figure}
\end{document}